\documentclass[10pt,A4,aps,pre,showpacs]{revtex4}
\usepackage{latexsym}
\usepackage{graphicx}
\usepackage{amsmath}

\begin{document}
\title{The marriage problem: from the bar of appointments to the agency}

\author{Alejandro Lage-Castellanos }
\affiliation{Henri-Poincar\'e Group of Complex Systems, 
Physics Faculty, University of Havana, La Habana, CP 10400, Cuba}
\affiliation{
Department of Theoretical Physics, Physics Faculty, University of 
Havana, La Habana, CP 10400, Cuba}
\author{Roberto Mulet}
\affiliation{Henri-Poincar\'e Group of Complex Systems, 
Physics Faculty, University of Havana, La Habana, CP 10400, Cuba}
\affiliation{
Department of Theoretical Physics, Physics Faculty, University of 
Havana, La Habana, CP 10400, Cuba}

\date{\today}



\begin{abstract}
We study the stable marriage problem from different points of
view. We proposed a microscopic dynamic that lead the system to a
stationary state that we are able to characterize analytically. 
Then, we derive a thermodynamical description of the Nash equilibrium
states of the system  that agree very well with the results of Monte Carlo
simulations. Finally, through large scale numerical simulations we
compare the Global Optimum of the society with the stable marriage of
lower energy. We showed that both states are strongly correlated 
and that the selffish attitude  results in a
benefit for most of the practitioners belonging to blocking pairs in
the Global Optimum of the society.
\end{abstract}
\pacs{05.20.-y, 01.75.+m, 02.50.Le}

\maketitle
\section{Introduction}

The Stable Marriage Problem (SMP) \cite{Gus89} describes a system where two classes of
agents (e.g. men and women) have to be matched pairwise. To each player
is assigned a list of preferred partners and the aim of the
problem is to find those states that are stable with respect to single
agent decisions, i.e. those states that are Nash equilibria \cite{Gib92}.

A Nash equilibrium state is a state in which,
being the
strategies of the other players constant,
 any variation of an agent's
strategy results in a worse performance for him
The concept
has become a fruitful source of inspiration for physicists,
 and many ``games'' have
been studied in this context. At variance with usual optimization
problems, where the  task is to maximize a global function, in
 Game Theory the main goal is to maximize the utility of the
agents of the system. 
In this context, the marriage problem has a parallel optimization
analogous, the Bipartite Weighted Matching Problem (BWMP)\cite{Mez85}, that is
also usually known as the Assignment Problem. In the
language of the marriage the BWMP consists in the determination of the state
that maximizes the global happiness of the society, or alternatively
the state that minimizes the unhappiness.  

Putting it in
other words,  given two sets of
$N$ agents, each one with a preference list for the possible
partners, two kind of problems are well posed: a) to find the assignment that maximizes the happiness of the
society (BWMP), b) to find those assignments which are stable with respect to the
individual decision of the agents (SMP).

From the technical point of view, both problems are described
assigning to each man (woman) an energy $x=X/N$ ($y=Y/n)$ if he(she) is married with a
woman(man) ranked $X$ ($Y$) in their list of preferences. The energy of
the system is therefore calculated as:

\begin{equation}
E=\sum_{i=1}^N x_i + \sum_{j=1}^N y_j
\label{eq:energy}
\end{equation}

\noindent and the BWMP problem is reduced to the search of the
assignment $\Pi$ that minimizes $E$ while the Stable Marriage Problem to the
search of those states that are Nash equilibria. 

Both problems have algorithmic solutions in polynomial time, and the
general properties of these solutions have been well studied in recent
years\cite{Ome97,Dzi00}. Moreover, the marriage problem has being extended to more
realistic situations, usually imposing limited information to the
agents \cite{Lau03}, or assuming correlations in the list of
preferences \cite{Cal00}. However
many relevant questions are still open: How does a real system reach the
Nash states? How stable is a Nash state to external perturbations? 
Or how much similar are the Optimum Global state of the system and
a Nash state? This last question for example, may gives some hint about
the tendency of the agents of a system, optimally matched, to act for
self benefit.

From the algorithmic point of view the assignment problem is usually assumed as a model where a
matchmaker decides who to pair with to optimize the happiness of the
society. In contrast, the solution of the Marriage Problem is
considered  as the natural stable
state to which the society evolves assuming infinitely rational agents
that share all the information of the game. However,
 it must be
keep in mind that the reality is by far more
complex. The dynamics of interaction between men and women, employers
and employees, buyers and sellers, etc, in society and economy is only
rarely enclosed in these kind of algorithms. Usually the dynamics is
more rich and do not warranty a convergence to a Nash equilibrium.

Therefore our aim in this paper is threefold, first to introduce and
to study a
local microscopic dynamics for the marriage problem 
that leads to a stationary state, (not necessary a Nash state).
Second, to study the evolutionary dynamics, between Nash states,  of the marriage
problem and to enclose it in a convenient thermodynamical
formalism. And finally to study the transition from the Global Optimum
state, decided by a matrimonial agency (matchmaker) to a Nash state and vice-versa.

The remaining of the paper is organized as follows. In the next
section we introduce a microscopic dynamic for the marriage problem and
propose a mean field description of this dynamic that characterizes
very well the stationary state of the system. From this solution we
are able to derive a strategy to improve the distribution of the
happiness in the society. Then, in section \ref{thermodynamics} we
propose a thermodynamical description of the stationary states of the
marriage problem. We compare the analytical and numerical results with
computer simulations and show a perfect agreement. Then, in section
\ref{states} we compare the statistical properties of the best Nash
state with the global optimum of the system. Finally, the conclusions
are outlined.

\section{The bar of appointments}
\label{bar}

In this section we introduce a new dynamics to model the evolution of
two kind of agents in the society. Keeping the analogy of the marriage
problem, we consider $N$ men and $N$ women, each one with a list of
preferences. 
At $t=0$ the system is prepared in a random configuration. 
Then, a man, belonging to a couple (let
us call it, couple $A_o$), and a woman from a different couple, ($B_o$) go
to  a bar (the bar of appointments) and meet together. Then, 
if both prefer to stay together, than with their original
couples,  (i.e. they are
better ranking in their respective preference list than the current
spouse and husband), 
they form a new couple $A_f$. Then, the abandoned partners, because of
the lonely form a new couple too $B_f$. The process is repeated and
at each time step, a man and a woman from different couples meet at
the bar and decide what to do.

Compared with the usual Marriage Problem this model may be
considered a limit situation where the preference lists of the
players are hidden to all, but the owner, a situation also closer to
the reality.
The model tries to be a simple cartoon of the dynamical behavior of many
economical and social systems. For example, in a completely free market job, a worker
have a meeting with a company. If the company sees that the new worker
may be more useful than some other already employed, and at the
same time the worker realizes that the offer of the company improves his current
position, then, he will choose to resign and the company to
assume him substituting one of his employees. The fired employee , will go
to the company of the former agent, now with a vacant position, and will be immediately
assumed. 

One may wonder if this kind of dynamics may lead to a Nash equilibrium
situation. Note that, being this the case, the system will be blocked
forever.  In fact, the system will be in Nash equilibrium, if and only if, there
are not men and women that prefer to stay together than with their
actual partners. Therefore, the Nash state are stable points of this
dynamic. 

\begin{figure}[!htb]
\includegraphics[angle=270,width=0.9\textwidth]{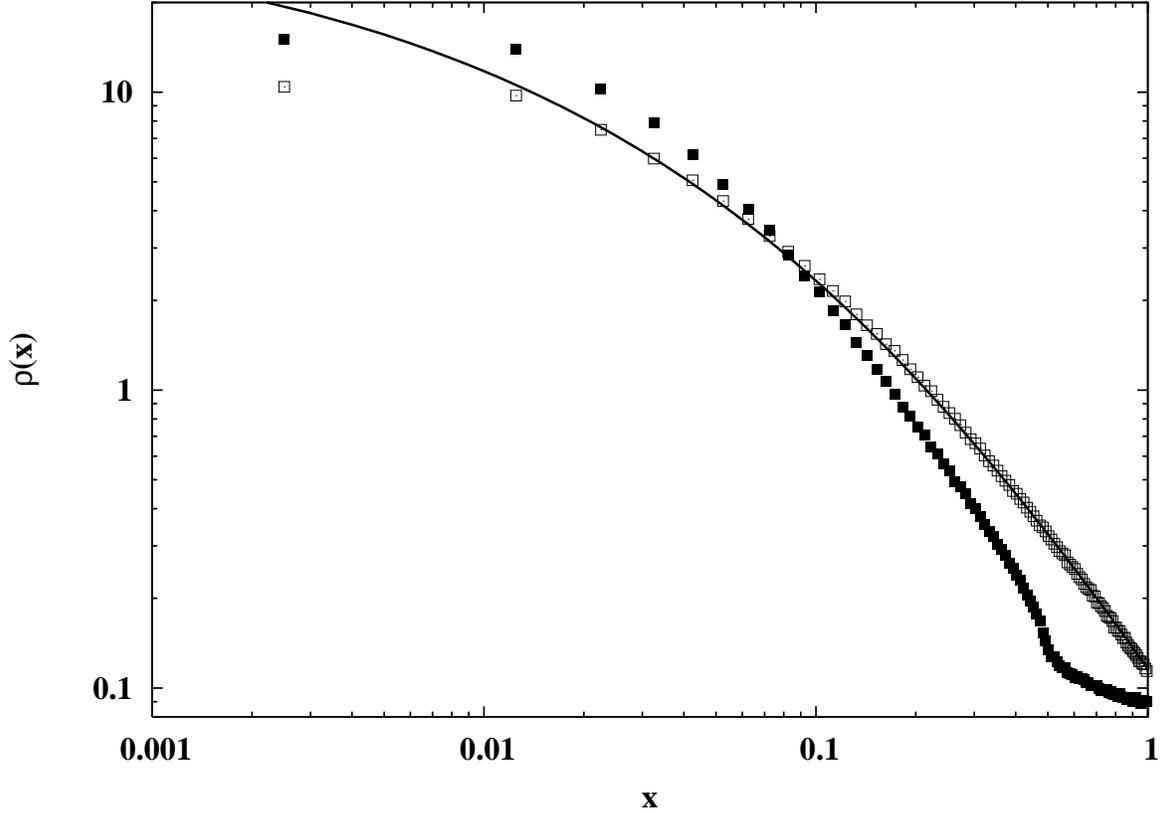}
\caption[0]{Stationary energy distribution for the bar of appointments
  (open symbols) and 
  the associated smarter strategy $N=500$ (close symbols). Symbols indentify
 simulation results for $N=500$ averaged over $1000$ instances. The
 continuos line represent the analytical solution of the model.
}

\label{fig:dist-bar}
\end{figure}

In figure \ref{fig:dist-bar} we plot the
probability density $\rho(x)$ obtained from the simulations that
 in the steady state a man(woman) has energy $x$ ($y$).

To analytically study this stationary state, we develop a mean field
description of this dynamic. At each time step, the man (woman)
that go to the bar, meet a woman (man) that is randomly located in
his (her) preference list and this location is independent of the person
met. In other words, at variance with the model described above where
the preference lists are fixed at the beginning of the simulation, now
the lists change every time that the agents go out from the bar. 

Under this new assumptions, we are able to determine which is the
probability $\rho(x,y)$ to find a couple where a man and a woman have energies $x$ and $y$
respectively. As usual,  a man has energy $x=X/N$ if he is
married with the woman ranked $X$ in his list of preferences, and
similarly  a woman has energy $y=Y/N$ if she is married with a man ranked
$Y$ in her list. The master equation of the stochastic process described above may
be written:

\begin{equation}
\frac{\partial \rho(x,y)}{\partial t}=\int_{(a,b)=0}^1 (P_{(a,b)
\rightarrow (x,y)} \rho(a,b) -  (P_{(x,y)\rightarrow (a,b)} \rho(x,y) )
da db
\label{eq:master}
\end{equation}

\noindent where $P_{(a,b) \rightarrow (x,y)}$ is the probability per
unit time that a man who belongs to a couple with energies $(a,b)$
goes to a new couple with energies $(x,y)$ after an appointment in the bar.

The first couple may be involved in a succeful appointment in two ways,
either the man of the couple meets a new woman preferred to the one he
is married with, or the woman meets a new man preferred to her current
husband. Keeping track of the different ways in which this may happen
(see appendix \ref{appendice}), the equation (\ref{eq:master}) may be written
as:

\begin{equation}
\frac{\partial \rho(x,y)}{\partial t}= \int_0^1 \int_x^1 \rho(a,b) da
db \int_0^1 \int_y^1 \rho(a,b) da db + <x>^2 -<x> \rho(x,y) (x+y)
\label{eq:master-dev}
\end{equation}

\noindent where $<x>$ is the mean energy of the men, and we used the
fact that because of the symmetry of the problem $<x>=<y>$.

To study the stationary solution of the problem, we make the right
hand side of (\ref{eq:master-dev}) equal to zero and obtain a closed, 
non-linear integral equation for $\rho(x,y)$.

\begin{equation}
\rho(x,y)= \frac{\int_0^1 \int_x^1 \rho(a,b) da
db \int_0^1 \int_y^1 \rho(a,b) da db}{<x>  (x+y)}
\label{eq:master-equil}
\end{equation}

Then, defining:

\begin{equation}
F(x)=\int_0^1 \int_x^1 \rho(a,b) da db
\label{eq:def-F}
\end{equation}

\noindent the probability density of having a couple with energies
$(x,y)$ becomes $\rho(x,y)=\frac{F(x) F(y) +<x>^2}{<x>(x+y)}$ and
substituting it in equation (\ref{eq:def-F}) we get:

\begin{equation}
F(x)=\int_0^1 \int_x^1 \frac{F(x) F(y) +<x>^2}{<x>(a+b)}
\label{eq:F}
\end{equation}

\noindent which can be further simplified for numerical porpoises
using another auxiliary function: 

\begin{equation}
T(a)=\int_0^1 db \frac{1}{<x>}
\frac{F(b)}{a+b}
\label{eq:T}
\end{equation}

\noindent such that:

\begin{equation}
F(x)=\int_x^1 da T(a) F(a) -  (x log(1+1/x) +log(1+x) - log 4) \frac{1}{x}
\label{eq:F-sol}
\end{equation}

In this way, we obtain two coupled equations,  one for $T(x)$
(\ref{eq:T}) and
another for $F(x)$ (\ref{eq:F-sol}) that can be solved in linear time. From $F(x)$ we
can trace back $\rho(x,y)$, and integrating with respect to one of the
variables, $\rho(x)=\int_0^1 \rho(x,y) dy$ the distribution of the
energies of the men (or women) in the stationary state. 

The solution
for $\rho(x)$ is presented with a continuous line in figure 
 \ref{fig:dist-bar}. The perfect coincidence with the simulations,
 shows  not only that the system is unable to reach a Nash
 equilibrium, but also that despite of their existence, and the fact that
 they are stable states of the simulation, they do not affect the dynamic
 proposed.

From the exact solution, we may obtain other properties of the
distribution, for example, the mean energy of the men.

\begin{equation}
X= N <x> = N \int_0^1 x \rho(x) dx \approx 0.167 N
\label{eq:ene-men}
\end{equation}

\noindent in excellent agreement with the simulation. 

This
result should be also compared with the solution of the problem obtained by
optimization algorithms. While with the dynamics proposed above, 
the energy growths linearly with $N$ (\ref{eq:ene-men}) in the original assignment and
marriage problems the energies growth as $N^{1/2}$. This impose, of
course not only a quantitative difference in the type of states
reached, Nash equilibria for the marriage, global equilibrium
for the assignment, and a global stationary state for our dynamics, but
also a qualitative difference between the previous two approaches and
our dynamics. This may suggest that a real society, where the
information of the agents is very limited, has less
capabilities for optimization than previously thought by the analysis
of the former problems.

Another conclusion that may be easily drawn from the distribution of
energies of the couples concerns the balance of energies between its
members. Writing $\rho(x,y)$ as a function of $\xi=x+y$, the energy
of the couple, and $\delta=x-y$, the energetic unbalance of the couple, it is
evident that $\rho(\xi,\delta)$ will be pair with respect to $\delta$
and therefore it will have an extreme at $\delta=0$. 

\begin{figure}[!htb]
\includegraphics[angle=270,width=0.8\textwidth]{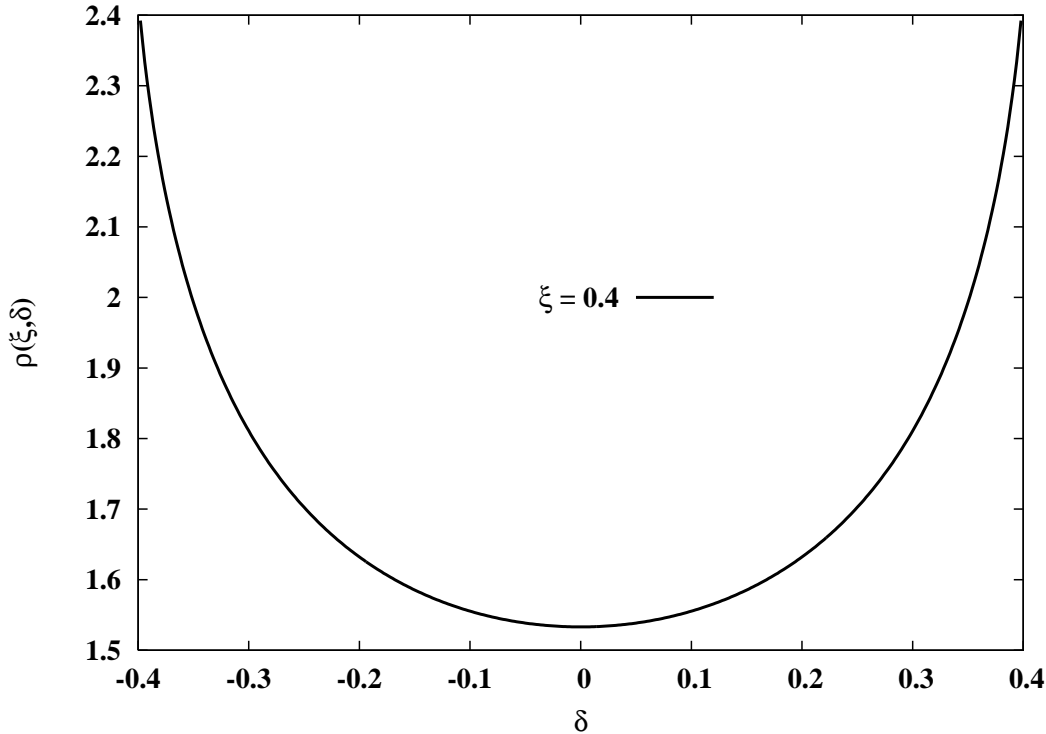}
\caption[0]{Probability to find a copy with energy $\xi$ as the
  function of the unbalance $\delta$}
\label{fig:unbalance}
\end{figure}

By numerical
inspection it is shown (see figure \ref{fig:unbalance}) that for small
values of the energies $\rho(\chi,\delta)$ has a minimum. This means that
couples with small energies (i.e. optimized), are very unbalanced,
if one of the members is very happy, the other is unsatisfied. This is
the price that the agents pay for acting in self-interest. They will spend half
of their life being the happiest of the relation, sharing with a partner that they  appreciate, while the other
half of their life they will be very unsatisfied.

Following the previous reasoning, we may imagine players that
adopt a smarter strategy when visiting the bar. They just need to have
a little more information. A man(woman), needs to know not only his
preference list but the one of the woman(men) he or she encounters in
the bar. Now, a man visiting the bar, will choose a new couple, not
only because she is better ranked that his actual partner, but because she
may guarantee a long term satisfaction. 

Let us, for simplicity, describe first this strategy for a man. 
Imagine a man with energy $x$, in average if his wife has energy $y$ he
will be abandoned after a time $1/y$. Once he is abandoned he will next
marry with a woman with random energy, in average $0.5$. Therefore, if
he has to decide between two women $w$ and $w'$ with energies $y>y'$  
respectively he follows the next reasoning: If
I marry the woman $w'$ I will be with her a time $1/y'$ being unhappy
a quantity $x'$, while to marry the woman $w$ will make me
unhappy a quantity $x$ for a time $1/y$ and a quantity $0.5$ for a
time $1/y'-1/y$. Then, a man will choose the woman $w'$ only if she
minimizes his unhappiness during this time, i.e. if:

\begin{equation}
\frac{x'}{y'}<  \frac{x}{y} +0.5 (\frac{1}{y'}-\frac{1}{y}) 
\label{eq:smart-choice}
\end{equation}

\noindent or, equivalently:

\begin{equation}
\frac{x'-0.5}{y'}<  \frac{x-0.5}{y} 
\label{eq:smart-choice-arran}
\end{equation}

Since the equation (\ref{eq:smart-choice-arran}) is symmetric with
respect to $w$ and $w'$ the man will choose the woman that minimizes
$\frac{x-0.5}{y}$. A woman, of course will behave in the same way.

Therefore when a man and a woman
meet in a bar, a man will leave her wife only if the value of
$\frac{x-0.5}{y}$ with the woman found at the bar is lower than the
same fraction calculated with his actual wife. A woman will do the same,
but substituting $x$ by $y$ in equation (\ref{eq:smart-choice-arran}).

The results of the simulations using this strategy appear in figure 
\ref{fig:dist-bar} with black symbols. As can be seen this new
strategy results in a general improvement of the energies of the
agents. The new distribution is higher for low values
of $x$ and smaller for large values of $x$ than the previous
one. The clear change in the slope of the distribution for
high energies shown in the figure suggests that the probability to
find a man or a woman with energy close to one is very low.

Calculating the mean energy of the men (or women) we find that it is
 $6\%$
 lower than the energy estimated with the first strategy. It must
also be noted that this improvement was obtained without the necessity to
increased considerably the capacity of the agents to get or to process the 
information and at the same time keeping their self-interest. 

From the market labor place, this
suggest that a good strategy for a job seeker may be not to look only for the must
profitable job, but also for a position where he is useful, such that,
in the short run, 
the employer may find difficult to find a good substitute for him.

\section{Thermodynamics of the stable states}
\label{thermodynamics}

In this section, we propose a thermodynamical formalism to
characterize the behavior of the system, assuming that it moves dynamically 
between the different Nash equilibria of the game. Compared with the
previous model, now, the agents,  share the full information of the
system and are able
to select their best strategy and to reach a Nash equilibrium state. 

At $t=0$, men and women will be in a Nash equilibrium state, and some
perturbation, that we will associate later with a temperature, kick
out the system from this state. However, since the agents
are infinitely rational, they immediately rearrange in a new 
Nash equilibrium state, not necessary the previous one.

A simple dynamics of this kind of behavior can be studied through
standard Monte Carlo techniques. Given the agents and their preference
lists, we first determine the different Nash equilibrium states of the
system. Then at $t=0$ we select one at random, and through a Markov
process we visit all the stable states,  with the following
probability, 
$p_{o\rightarrow f}= min[1,exp(-\beta (E_f-E_o))]$, where $E_f$ and
$E_o$ are the energies of the final and initial states of the system. 

The
temperature of the system $1/\beta$, represents the external drive
received by the agents to change their equilibrium state. A low value
of $\beta$ reflects a situation where the agents, despite the fact
that they are in a Nash equilibrium, tend to change frequently their
partners, looking for  better matchings. A large value of $\beta$
reflects the opposite, the tendency of the agents to explore new
possibilities is strongly reduce.

\begin{figure}[!htb]
\includegraphics[angle=270,width=1\textwidth]{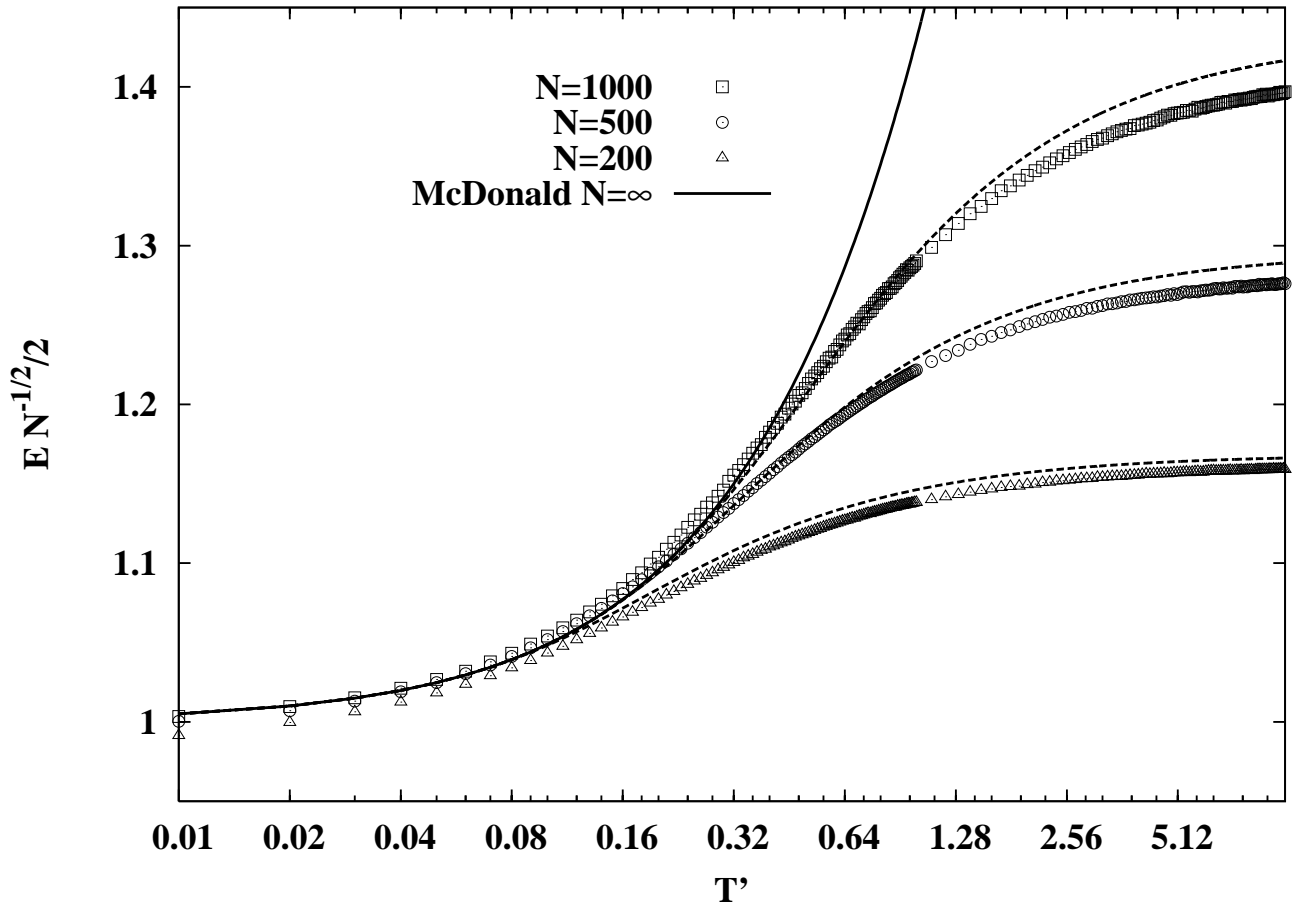}
\caption[0]{Energy of the system as a function of the temperature
  $T$. Every point is the average over $100$ instances during the simulation. The continuous
  line is the analytical result and the discontinuous one, represents
  the numerical calculations}
\label{fig:E-term}
\end{figure}

\begin{figure}[!htb]
\includegraphics[angle=270,width=1\textwidth]{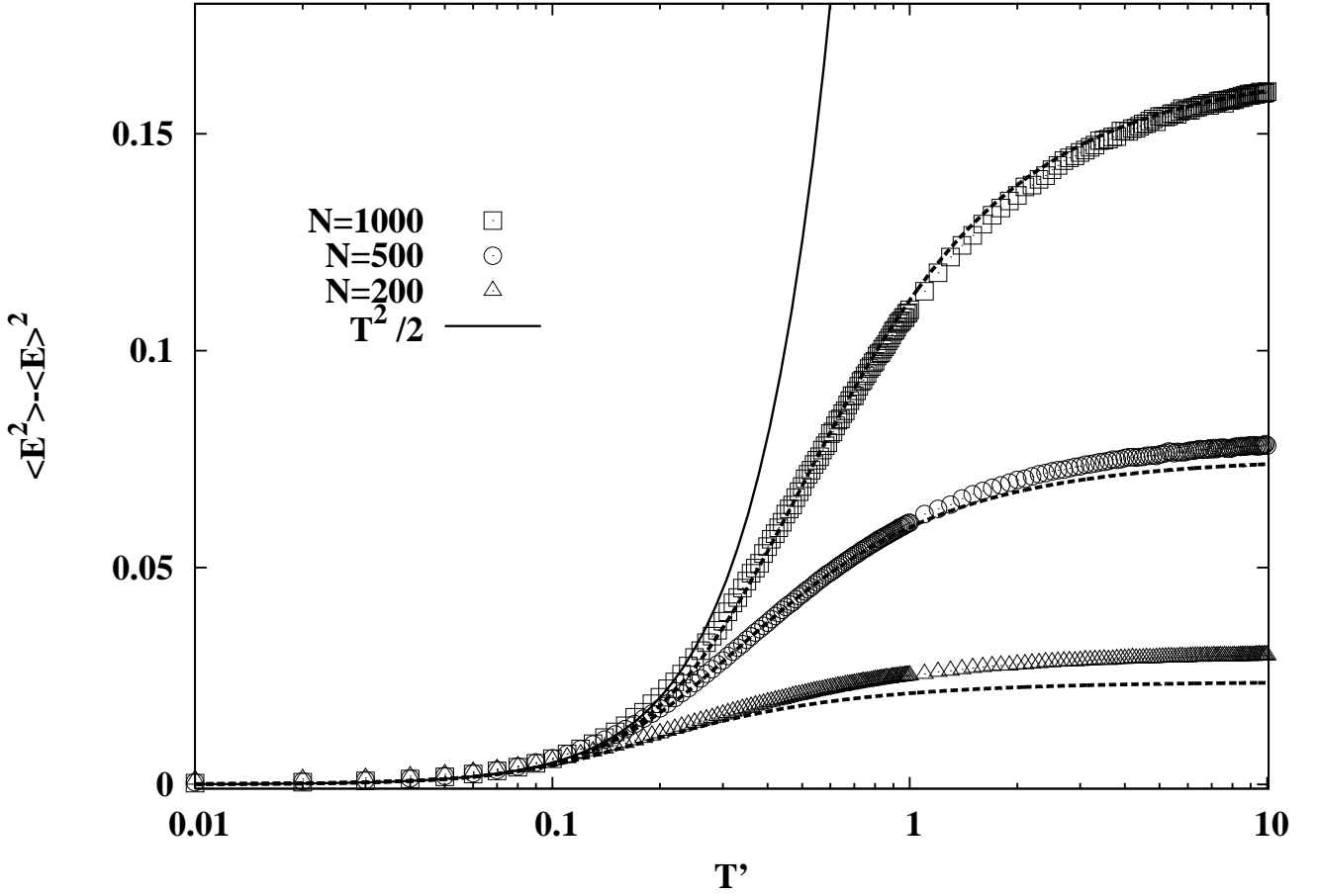}
\caption[0]{Specific heat of the system as a function of the temperature
  $T$. Every point is the average over $100$ instances during the simulation. The continuous
  line is the analytical result and the discontinuous one, represents
  the numerical calculations}
\label{fig:C-term}
\end{figure}

We calculate the total energy of the system, and its fluctuations as a
function of $T$. The results of the simulations appear with symbols in figure
\ref{fig:E-term} and \ref{fig:C-term}. Note that both the total energy 
 and the fluctuations of the system decrease with the temperature. For
 large temperatures, however, the energy saturates, a finite size
 effect that will be discussed below.

To explain these results, we perform a thermodynamical analysis based on
the approach developed by Dzierzawa et al\cite{Dzi00}. to study the number of
equilibrium states of the system.

The number of stable (Nash) states in the system may be written as:

\begin{eqnarray}
S &=&\int_{}^{}dX \int_{}^{}dY\rho(X)\rho(Y)e^{-XY} \label{eq:IntNSS}
\label{eq:nss}
\end{eqnarray}

\noindent where $X=\sum_{i=1}^N x_i$, $Y=\sum_{i=1}^N y_i$, and
$\rho(X)=\frac{X^{N-1}}{\Gamma(N)}\left(1-\exp{(-\frac{X}{N})}\right)^N$. 

From equation (\ref{eq:nss}) and, the known result that the density of
states for $N$ 
large enough is concentrated around the curve $XY=N$\cite{Ome97} we
may calculate the density of states of the system:

\begin{eqnarray}
D(E)&=&\int_0^N dX \int_0^N dY \rho(X)\rho(Y)e^{-XY} \delta(XY-N) \delta(X+Y-E) \nonumber \\
    &\approx&\frac{1}{\sqrt{E^2-4N}}   
\label{eq:DE}
\end{eqnarray}

\noindent valid in the interval $(2 \sqrt{N}, logN+N/log N)$. Then, to get some
insight on the thermodynamical behavior of the system we calculate,
using (\ref{eq:DE}), the partition function of the system at a
finite value of $\beta$:

\begin{eqnarray}
Z(\beta)&=&\int_{2\sqrt{N}}^{N/\log{N}} dE D(E) e^{-\beta E} \nonumber \\
&=&\int_{2\sqrt{N}}^{N/\log{N}} dE \frac{e^{-\beta E}}{\sqrt{E^2-4N}}
\label{eq:Zb}
\end{eqnarray} 

\noindent which in the new variable $E'=\frac{E}{2 \sqrt{N}}$ may be
written as:

\begin{eqnarray}     
   Z(\beta)&=&   \int_{1}^{\sqrt{N}/(2\log{N})} dE' \frac{e^{-2\sqrt{N}\beta E'}}{\sqrt{E'^2-1}}
\label{eq:Zb'}
\end{eqnarray}

\noindent that can, in turn, be analytically solved in the thermodynamic limit $N
\rightarrow \infty$,

\begin{eqnarray}
Z(\beta) &\simeq& \int_{1}^{\infty} dE' \frac{e^{-2\sqrt{N}\beta E'}}{\sqrt{E'^2-1}} \nonumber \\
 &=& K_0(2\sqrt{N}\beta) 
\label{eq:sol_Zb}
\end{eqnarray}

\noindent where $K_0(x)$ is the MacDonald function of order zero. From
 (\ref{eq:sol_Zb}) it is an easy exercise to derivate the
 thermodynamical variables of the system. The energy follows \cite{pepe}:

\begin{eqnarray}
E'(\beta')&=&  \frac{K(1,\beta')}{K(0,\beta')} 
\label{eq:EvsT}
\end{eqnarray}

\noindent and the specific heat \cite{pepe}:

\begin{eqnarray}
C'(\beta'=1/T')
&=&\beta'^2\left(\frac{1}{2}+\frac{K(2,\beta')K(0,\beta')-2K(1,\beta')^2}{2
    K(0,\beta')^2}\right) 
\label{eq:CvvsT}
\end{eqnarray}

\noindent where $\beta'=2 \sqrt{N}\beta$. 

These results (equations (\ref{eq:EvsT}) and (\ref{eq:CvvsT}) appear as solid lines in figures \ref{fig:E-term} and
\ref{fig:C-term}. Note that the agreement with the simulation 
is very good at low temperatures, but a clear discrepancy exists 
for large values of $T$.

\begin{figure}[!htb]
\includegraphics[angle=270,width=1\textwidth]{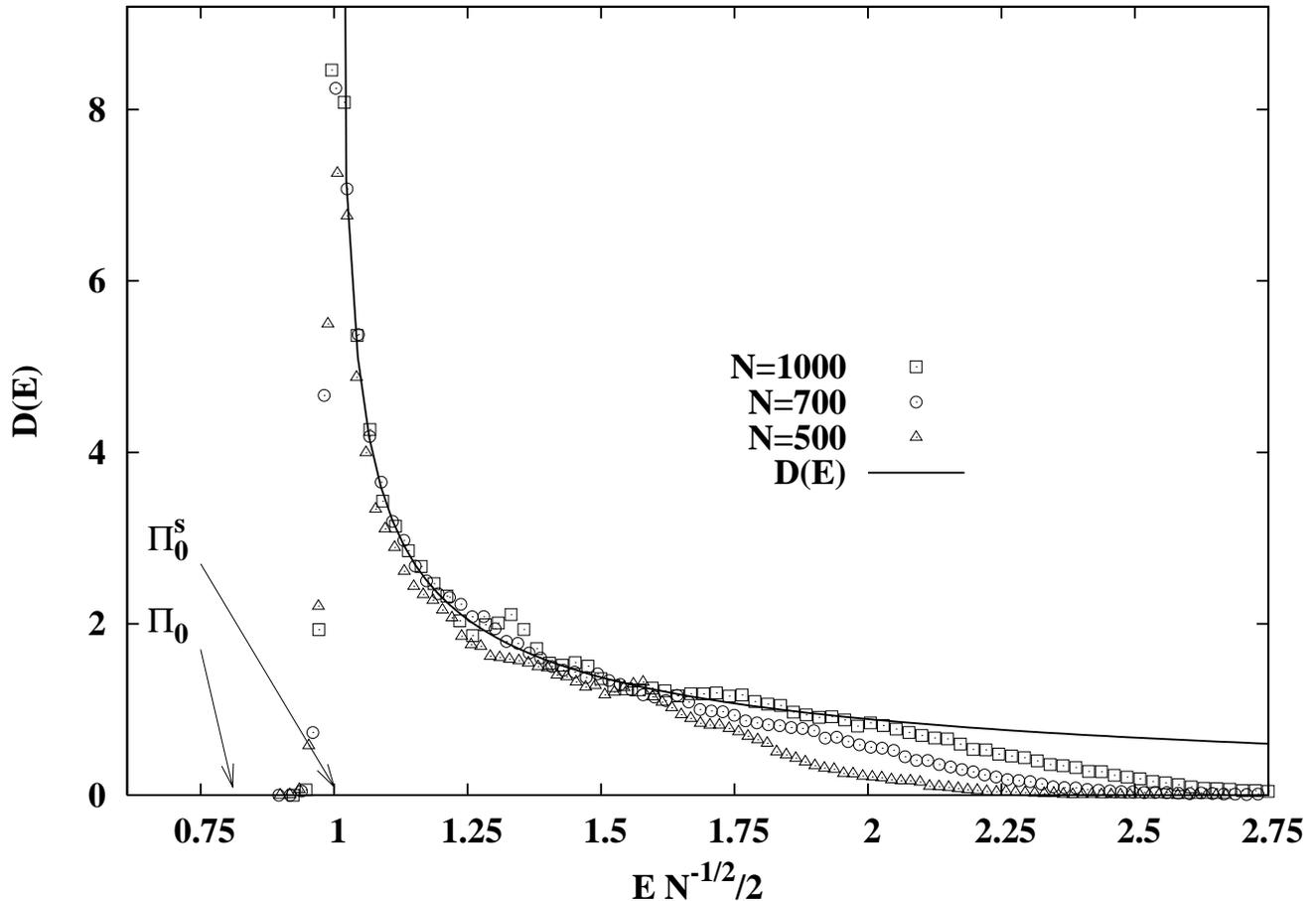}
\caption[0]{Histogram of the energies of the stable states averaged
  over $1000$ instances of the problem (symbols). The continuous line
  represent the analytical solution. (see eq. (\ref{eq:DE})}
\label{fig:DE}
\end{figure}

This discrepancy is due to the
approximation used for the calculation of $D(E)$ first, and latter of
$Z(\beta)$. In the calculation of the density of states we use the
fact that $XY=N$, a result that is true only for very large $N$. In
fact, the comparison of these analytical results (solid line) with the numerical
computation (symbols) of the density of states shown in figure \ref{fig:DE},
reflects that for large values of $E$ this approximation is not longer
true. Moreover, taking the thermodynamic limit in equation (\ref{eq:Zb'})
to get the analytical solution (\ref{eq:sol_Zb}) for $Z(\beta)$ , is not mathematically
justified for $\beta \rightarrow 0$. 

A more precise, while not
analytical treatment of the problem may be obtained solving
numerically, for fixed values of $N$ the equation  (\ref{eq:Zb'}) and
the thermodynamical derived equations for $E$ and $C_v$. These
numerical results appear also in figures \ref{fig:E-term} and
\ref{fig:C-term} with dashed lines, showing  a much better coincidence
with the simulations.

Therefore, the comparison of the analytical results obtained above with the Monte Carlo
 simulations, show that the approach developed by Dzierzawa et al. \cite{Dzi00} can be extended to describe the dynamical
 behavior of  
these kind of systems. Moreover, the analysis of the results lead us
 to conclude
 that a too dynamical society (high temperature) will lead the agents to explore configurations that, while
 stable are, in average for the society, less satisfactory.

\section{Stable states versus global optimization}
\label{states}

It is also worth to compare which are the similarities and
differences between the stable states of the marriage problem and the global optimum of the
society. The comparison is helpful first because these problems
have been rarely studied together beyond the average energy analysis,
and to have a more general idea about which are really the differences
and similarities of both kind of solutions may give some insight on
the interplay between global and personal optimization.
Moreover, it can give also some clue on which is the best strategy that
should follow a matchmaker (or a matrimonial agency) when facing the
problem of pairing two subsets, that can a posteriori take their own decisions.

To make this comparison we performed extensive numerical
simulations and compare the statistical properties of the matching in
two states, the Global Optimum of the system $\Pi_0$ , calculated using the
Hungarian method\cite{Mez85}, and the Stable Marriage with lower
energy $\Pi_0^s$. 

It was already proved that the energy of both kind of problems is
different. For the BWMP $E_{B}=1.617 \sqrt{N}$\cite{Mez85,Ome97}
and for the best marriage $E_{M}=2 \sqrt{N}$\cite{Ome97}.
Moreover from figure \ref{fig:DE}, it becomes clear that there is
an energetic gap, between the solution of the Marriage Problem and the solution of the
Weighted Assignment Problem.

An interesting quantity to measure between both states is the
distance, defined as the number of different couples in the two matchings:

\begin{equation}
D(\Pi^a,\Pi^b)=\frac{1}{N} \sum^N_{i=1}(1-\delta_{\Pi_{i}^{a},\Pi_{i}^{b}} )
\label{eq:distancia}
\end{equation}

\noindent where $\Pi^x$ represents a matching. We as $\Pi^a$ and
$\Pi^b$ the configurations of the stable marriage with lower energy ($\Pi_0^s$)and
the Global Optimum ($\Pi_0$)respectively for the same set of preference lists.

\begin{figure}[!htb]
\includegraphics[angle=270,width=1\textwidth]{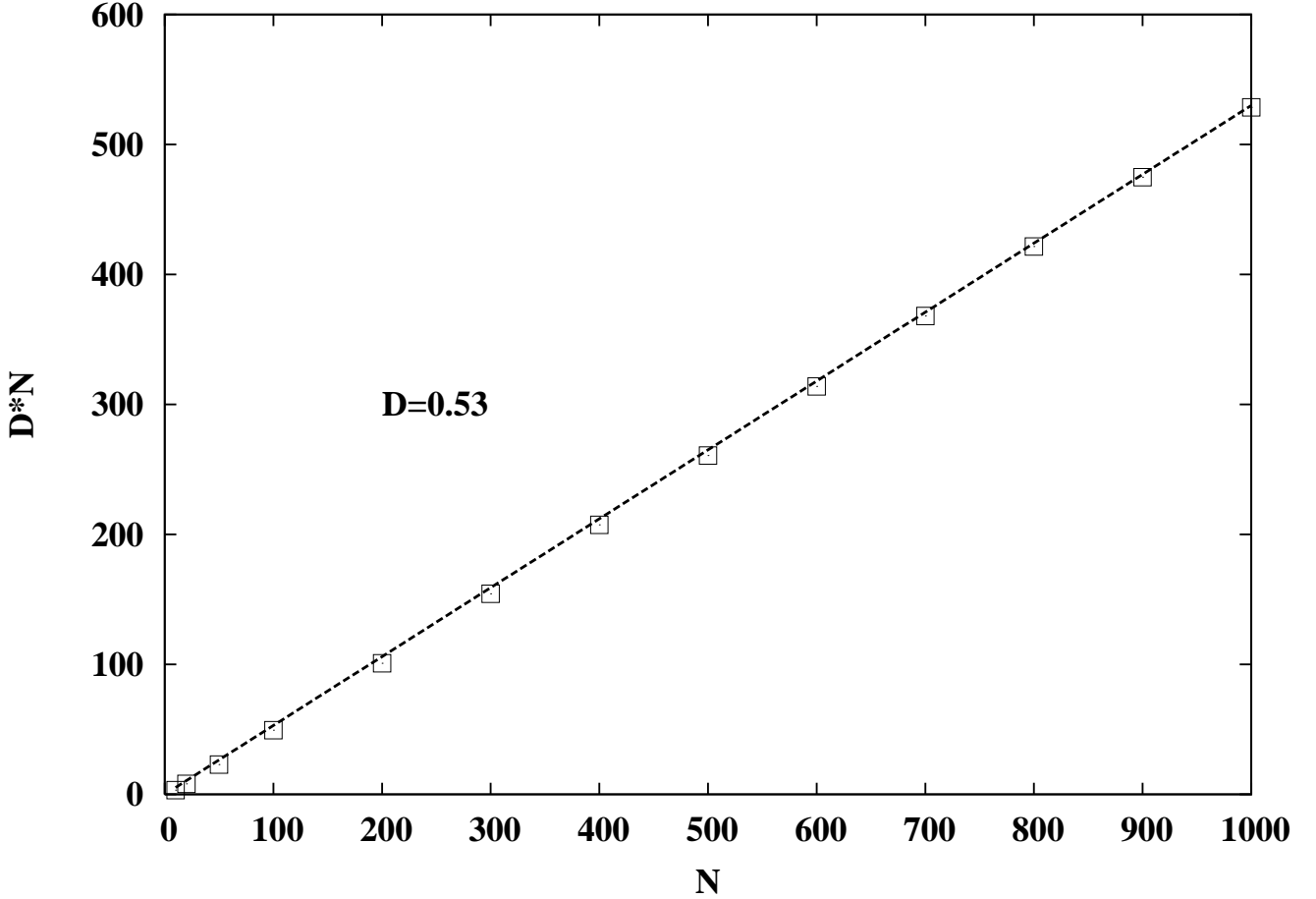}
\caption[0]{Distance $N\cdot D(\Pi_0,\Pi_{0}^s)$ between the Global
  Minimum of the system and the Optimum Stable State. For each value
  of $N$ the points reflect averages over $1000$ instances}
\label{fig:D-vs-N}
\end{figure}

Therefore, for a given set of preference lists of the players, we
calculate the Global Optimum of the society, and determine the Stable
Marriage State with lower energy (Optimum Stable State), then we calculate the distance between
these two matchings.
After averaging over $1000$ instances we obtain that $D=0.53$
independently of $N$ (see figure \ref{fig:D-vs-N}). This result must be
compared with the fact that the distance between two randomly taken
matching goes as  $1-1/N$, see appendix \ref{appendiceB}, that proves that there is
a strong correlation between the two states. Approximately half of the
couples are the same.

Moreover, we will like to know, from those couples that are different,
how many of them and in which extend they improve their situation when the
system moves from the Optimal Stable State to the Global Optimum. 
Again, averaging over $1000$ instances we find that the
$24\%$ of the players improve his situations while the $29\%$ get
worse matches. However, the improvement of the formers is large enough to
compensate the fact that more people get worst.

\begin{figure}[!htb]
\includegraphics[angle=270,width=0.9\textwidth]{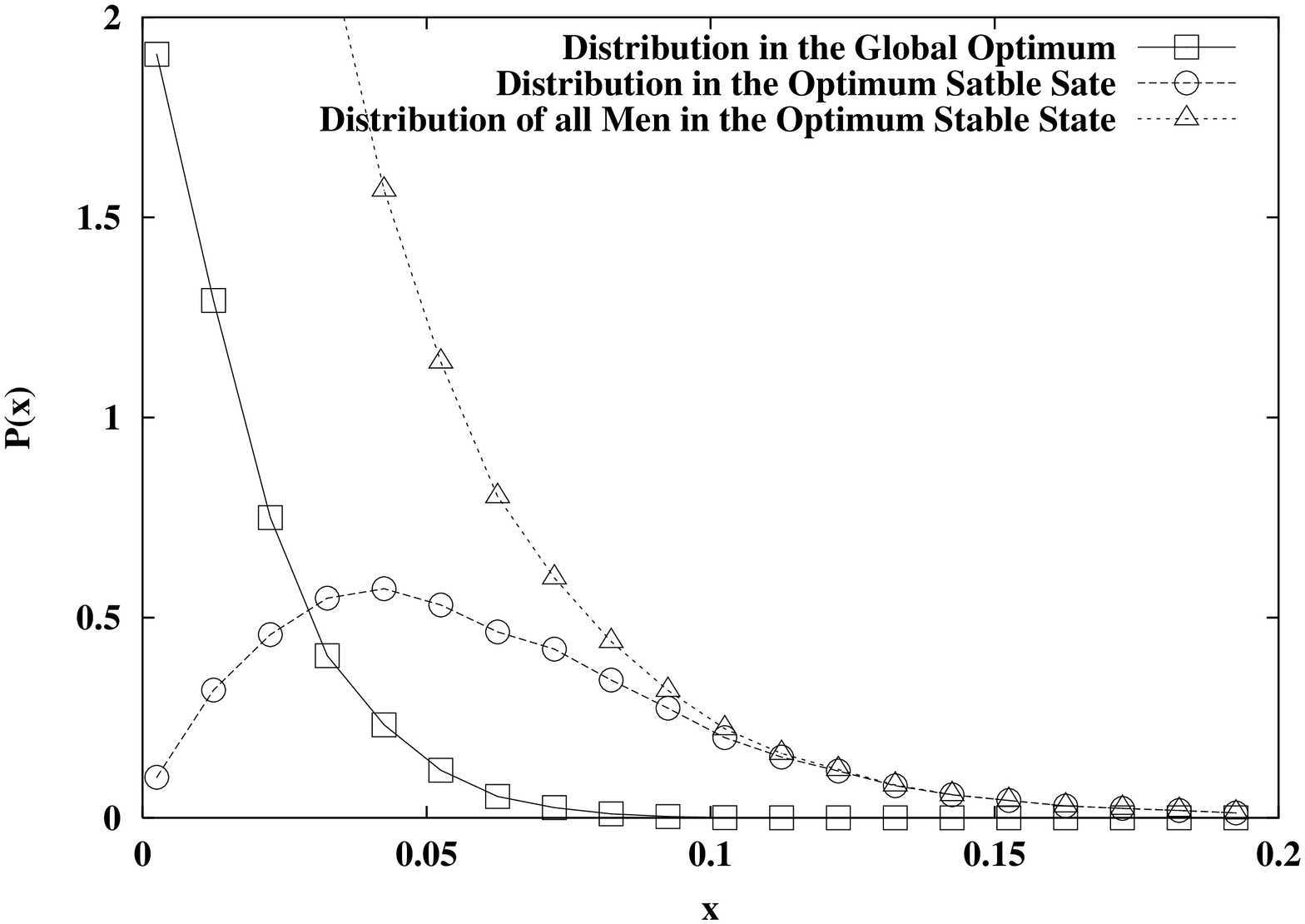}
\caption[0]{Distribution of energies of the men that benefit from the transition
 to the Global Minimum for the Optimal Stable State. Every point
 represents an average over $100$ instances of size $N=1000$}
\label{fig:OSS-GO}
\end{figure}

The results appear in figure \ref{fig:OSS-GO}.  The data points marked
by 
$\Box$ represent the histogram (measured when the system is in the
Global Optimum) of the energies of the men 
benefited in the transition from the Optimal Stable State to the
Global Optimum.
The points with $\bigcirc$ reflect the  histogram of the energies of the
same men but in the Optimum Stable State and those marked 
with $\bigtriangleup$ represent the histogram of all the men (not only
those benefited) in the Optimum Stable State. As can be seen from the figure, the tail of the energy distribution
of the men in the stable state disappears when the system
moves to the Global Optimum. It means that the transition to the
Global Optimum from the Optimum Stable State is achieved improving,
mainly, the situation of the people  matched worst in this state.

\begin{figure}[!htb]
\includegraphics[angle=270,width=0.9\textwidth]{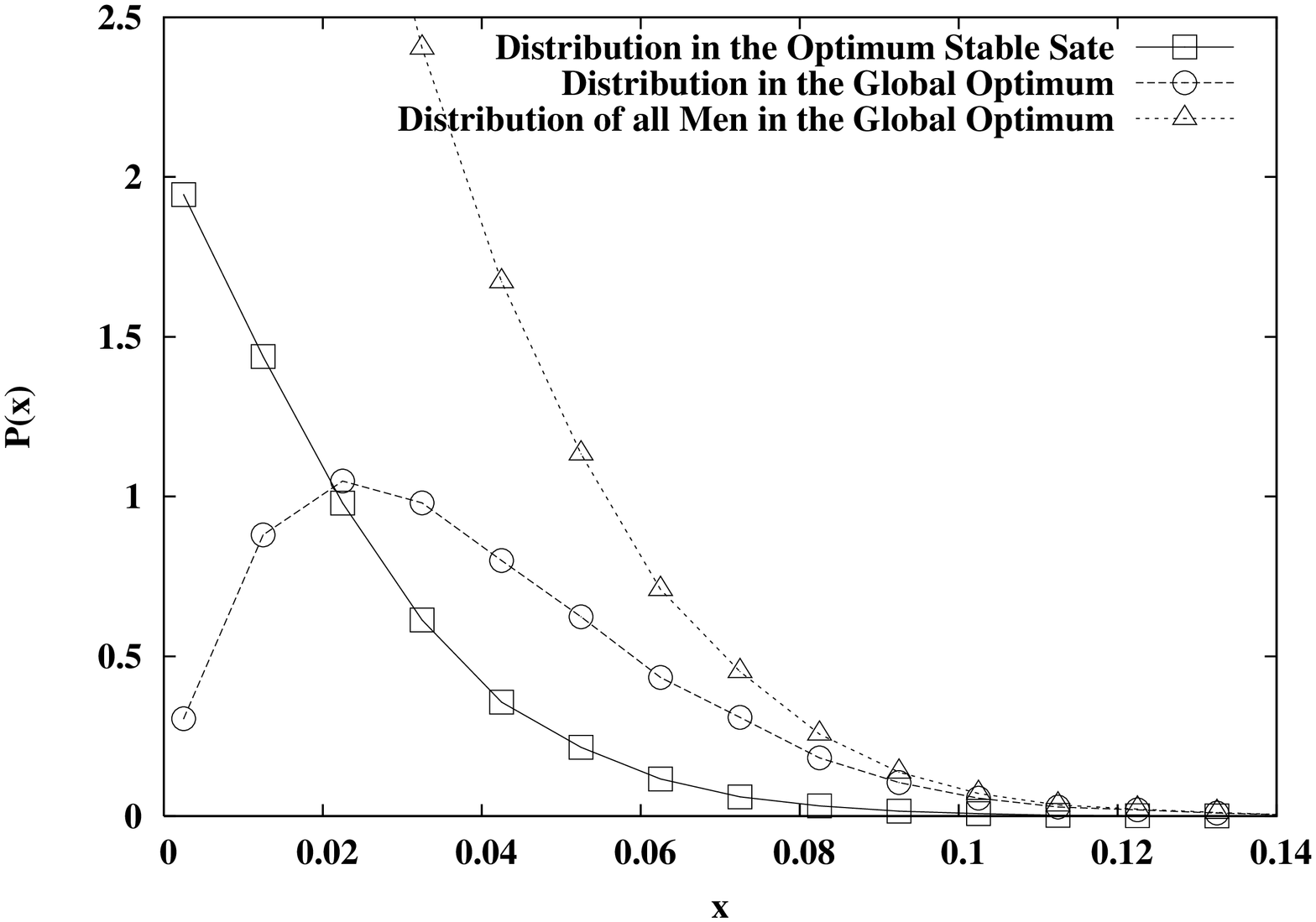}
\caption[0]{Distribution of energies of the men that benefit from the transition
 to the Optimum Stable State from the Global Minimum. Every point
 represents an average over $100$ instances of size $N=1000$}
\label{fig:GO-OSS}
\end{figure}

Analogously figure  \ref{fig:GO-OSS} reflects the opposite situation, in which a
society goes from the Global Optimum to the Optimum Stable State. The
points marked with $\Box$ represent the histogram of the energy of all
the men in the Optimum Stable State, while the other curves reflect the histogram of
the energies of the men benefited with the transition (measured in the
Optimum Stable State) ($\bigcirc$) and
the histogram of the energies of all the men in the Global Optimum 
($\bigtriangleup$). Again the tail of the distribution of the men in
the Global Optimum contains those players that are benefited in the
transition from the Global Optimum to the Optimum Stable State.

Then, the  main conclusion to be drawn from figures
\ref{fig:OSS-GO} and \ref{fig:GO-OSS} is that the players with
higher energies would be the most motivated to make a transition from one
state to the other (if they were informed about
their presumed situation in the other state).

Moreover, it is
interesting to find the number of unstable couples $N_{uc}$ in the Global
Optimum. This is a better measure of the instability of the system and
may give an idea of how many people will be tempted to act in
self-interest if the society is in the Global Optimum. 

\begin{figure}[!htb]
\includegraphics[angle=270,width=0.9\textwidth]{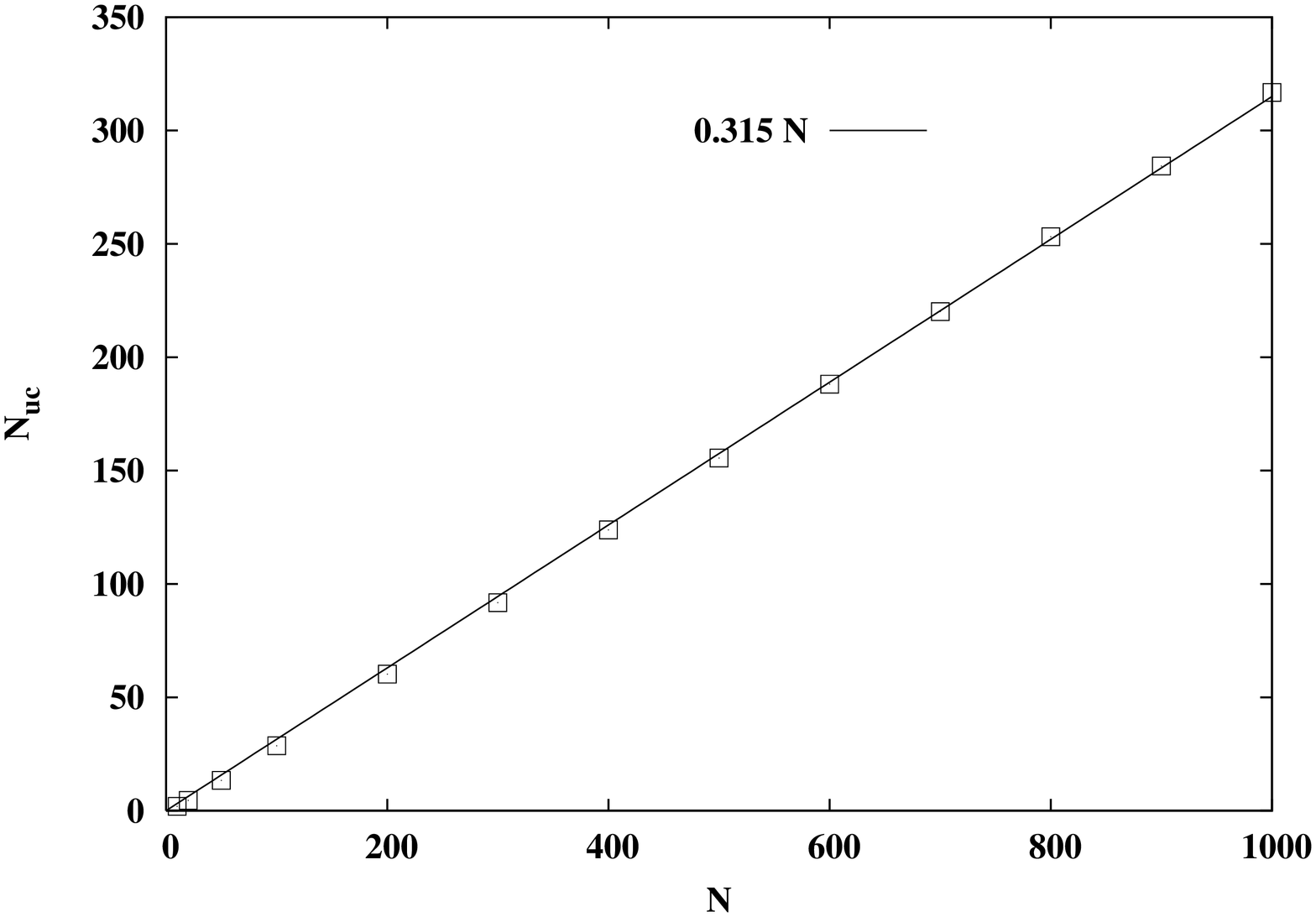}
\caption[0]{Number of unstable couples $N_{uc}$ as a function of the
  system size $N$. The points reflect average values over $1000$ instances}
\label{fig:NUC}
\end{figure}

Simple probabilistic arguments (see appendix \ref{appendiceC}) suggest
that $N_{uc}$ growths linearly with $N$. This is, indeed, the case (see figure \ref{fig:NUC}). 
Therefore, since $N_{uc}$ is a measure of how many couples are
interested in acting for self-interesting the 
system is in the Global Optimum, considering that $N_{uc} \sim N$
and that the number of possible couples growths with $N^2$ the ratio of the
couples to be forbidden to keep the system in the Global Optimum decreases as $1/N$. This mean that the
larger the society, the easy it is to keep it in the Global
Optimum. While difficult to compare with real situations (where it has
been never achieved the Global optimum) this result agrees with the standard
notion that the larger the system the more stable it is. This is
certainly also true in physical systems where the fluctuations decrease as $1/\sqrt{N}$. 

Finally, we ask, whether this perception of improvements felt by the
player implied in unstable couples was real. In other words, how many  of the
people perceiving that may get benefits acting for self-interest will
really find a better partner in the Nash state of lower
energy. Studying by numerical simulations $100$ systems of size $N=1000$ we concluded that the
percentage of the people that have this feeling and will really
improve his situation is nearly $80\%$. 

Summarizing, the results of this section suggest that, subject to
evolutionary selection the self-fish attitude will prevail over a
Global Optimum in a
competitive society. People that perceives that may improve their
situation acting for his self-interest will really improve it taking
the initiative and changing the partner.

\section{Conclusions}

We have studied the stable marriage problem from different point of
views. We first proved that it is not easy to invent a simple
microscopic dynamics
for agents with limited information that lead to a Nash equilibrium state. However,
we were able to characterize analytically one such dynamics that
converges to a stationary state. From these results we proposed for
the players a
strategy  that greatly improves their
performance. We then focus our attention on the stable states of the
marriage problem and derive a thermodynamical description of these
states that agree very well with the results of Monte Carlo
simulations. Finally, through large scale numerical comparison we
compare the Global Optimum of the society with the stable marriage of
lower energy. We showed that both state are strongly correlated and
more interesting that the selffish attitude will indeed result in a
benefit for most of the practitioners involved in unstable couples.

\section{Appendix A}
\label{appendice}

The master equation for $\rho(x,y)$
\begin{eqnarray}
\frac{\partial \rho}{\partial t}(x,y)=\int_{a,b\:=0}^{1} \displaystyle\left( P_{a,b\rightarrow x,y}\rho(a,b)- P_{x,y\rightarrow a,b} \rho(x,y)\displaystyle\right)da\:db \label{eq:anMaestra}
\end{eqnarray}

\noindent though very intuitive has the terms $P_{a,b\rightarrow x,y}$ which are
a little bizarre, and are not needed to do the calculation. The
integration can be done directly without calculating explicitly this
terms. It is suitable then to write the master equation as a balance
of probabilities like this:

\begin{eqnarray}
\frac{\partial \rho}{\partial t}(p_0)=W_{\rightarrow p_0}- W_{p_0  \rightarrow} \label{an:MaestraW}
\end{eqnarray}

\noindent where $p_0=(x_0,y_0)$ is the energy vector of a couple and $W_{\rightarrow p_0}$ is the probability of having a situation in which a couple $p_0$ is created after an effective appointment, while $W_{p_0 \rightarrow}$ is the probability of a situation in which such a couple finds itself involved in an effective appointment that breaks it apart. The position of the arrow stands for whether the effective appointment reaches or starts from a couple $p_0$.

In any appointment there are four couples to be considered: $p_1=(x_1,y_1)$ and $p_2=(x_2,y_2)$ which are the original couples, and $p_1'=(x_1',y_2')$ and $p_2'=(x_2',y_1')$ which are the possible new couples. Given that the primed values are taken randomly within $(0\ldots 1)$, the probability density of a given point of the space of probabilities is $\rho(p_1,p_2,p_1',p_2')=\rho(p_1)\rho(p_2)$, which is normalized. The situations in which a couple $p_0$ is obtained after an effective appointment happens both when  $p_1'=p_0$ and when $p_2'=p_0$. While the situation in which a couple $p_0$ is destroyed takes place when  $p_1=p_0$ and also when $p_2=p_0$. So the master equation can be written as:
\begin{eqnarray}
\frac{\partial \rho}{\partial t}(p_0)=W_{ p_1'= p_0}+W_{p_2'= p_0}- W_{p_1=p_0 }- W_{p_2=p_0 }  \label{an:MaestraW2}
\end{eqnarray}

The probability of any given situation $A$ can be measured by the integration of a function $A(p_1,p_2,p_1',p_2')$ that is equal to 1 in the points of the probability space in which the situation occurs while is 0 in all other points. An appointment will be effective if $x_1'<x_1$ and $y_2'<y_2$. Then we can measure the situations $A_p$ in which $p=p_0$ and an effective appointment occurs through the function
\begin{equation}
A_p=\theta(x_1-x_1')\theta(y_2-y_2')\delta(p-p_0)
\end{equation}
where $p$ can be either of the four couples involved in the appointment.

Any of the probability densities $W_p$ in equation (\ref{an:MaestraW2}) are given by:
\begin{equation}
W_{p=p_0}=\int_0^1 dp_1 dp_2 dp_1' dp_2' \rho(p_1)\rho(p_2) \theta(x_1-x_1')\theta(y_2-y_2')\delta(p-p_0) 
\end{equation}

This integration is made straightforward. To show it let us take the case of $W_{p_1'=p_0}$

\begin{eqnarray}
W_{p_1'=p_0}&=&\displaystyle \int_0^1 dp_1 dp_2 dp_1' dp_2' \rho(p_1)\rho(p_2) \theta(x_1-x_1')\theta(y_2-y_2')\delta(p_1'-p_0) \nonumber \\
&=&\displaystyle \int_0^1 dp_1 dx_1' dy_1' \rho(p_1)\theta(x_1-x_1')\delta(x_1'-x_0)\:\: \int_0^1 dp_2 dx_2' dy_2' \rho(p_2)\theta(y_2-y_2')\delta(y_2'-y_0) \nonumber \\
&=&\int_0^1 dp_1 \rho(p_1)\theta(x_1-x_0) \:\:\int_0^1 dp_2 \rho(p_2)\theta(y_2-y_0) \nonumber \\
&=& \int_{x_0}^1 \int_0^1 \rho(x,y)dxdy \:\: \int_0^1\int_{y_0}^1
\rho(x,y)dxdy
\end{eqnarray}

Doing the same with the rest three terms of the master equation (\ref{an:MaestraW2}) we easily get:
\begin{eqnarray}
\frac{\partial \rho}{\partial{t}}(x,y) &=&\displaystyle \int_{0}^{1}\int_{x}^{1} \rho(a,b) da\:db 
                                       \int_{y}^{1}\int_{0}^{1} \rho(a,b) da\:db \nonumber + <x><y>\\
                                       && -x<y>\rho(x,y) -y<x>\rho(x,y) 
\end{eqnarray}

\section*{Appendix B}
\label{appendiceC}

There is a bijection between states of a system of size $N$ and the permutations of the numbers $(1,2\ldots N)$ which results obvious if we consider that a given permutation $\Pi$, stands for the state with couples $\{(1,\Pi_1),(2,\Pi_2)..(N,\Pi_N)\}$. Two random states are given by two random permutations $\Pi^a$ and $\Pi^b$.

Without lost of generality we can take one of this states fixed $\Pi^a=(1,2\ldots N)$ while leaving all the randomness to the other. Now asking the distance is equivalent to ask how many of the numbers coincide with their position in the random permutation:
\begin{eqnarray}
\Pi&=&4\:\:2\:\:5\:\:3\:\:1\:\:2  \nonumber \\
&&1\:\:2\:\:3\:\:4\:\:5\:\:6 \nonumber
\end{eqnarray} 

Lets say $P(k,N)$ is the probability that $k$ of $N$ numbers will coincide with their position in the random permutation. There are $(^k_N)$ ways to pickup such $k$ numbers, and for each of these there are $(N-k)!$ ways to order the rest $N-k$ numbers, of which a fraction $P(0,N-k)\simeq P(0,N)$ is such that no other number ends at its position, thus ensuring that only the first $k$ numbers do. Then we have a relation between $P(k,N)$ and $P(0,N)$:
\begin{eqnarray}
P(k)&=&\frac{(_{N}^k)\:(N-k)!\:P(0)}{N!} \nonumber \\
    &=&\frac{P(0)}{k!}  \\
\end{eqnarray}
$P(0)$ can be calculated by the inclusion exclusion principle as:
\begin{eqnarray}
P(0) &=& \sum_{q=0}^{q=N} \frac{(-1)^{q+1}}{q!} \nonumber \\
     &\simeq& e^{-1} \\
\end{eqnarray}
Then the mean value of $k$ is:
\begin{eqnarray}
\overline{k} &=& \sum_{k=0}^{k=N} k P(k) 
             = \frac{1}{e} \sum_{k=0}^{k=N} \frac{k}{k!} 
             = \frac{1}{e} \sum_{k=0}^{k=N-1} \frac{1}{k!}
             = \frac{1}{e}\:e \nonumber \\
             &=& 1 
\end{eqnarray}
and the distance between random states is in average:
\begin{eqnarray} 
D(\Pi^a,\Pi^b)&=&\frac{1}{N} \sum^N_{i=1}(1-\delta_{\Pi_{i}^{a},\Pi_{i}^{b}} ) \nonumber \\
&=&\frac{N}{N}-\frac{\sum^N_{i=1}\delta_{\Pi_{i}^{a},\Pi_{i}^{b}}}{N} \nonumber \\
&=&1-\frac{\overline{k}}{N} \nonumber \\
&=&1-\frac{1}{N} \:\: \sim 1 \nonumber
\end{eqnarray}

\section*{Appendix C}
\label{appendiceB}

A mean field approximation to the number of unstable couples $N_{uc}$ can be work out considering that a man with energy $x$ and a woman with energy $y$ form an unstable couple with probability $xy$. Then, in average any two persons $m$ and $w$ will form an unstable couple with probability:

\begin{eqnarray}
P_{h,m} & =&\int_{0}^{1}\int_{0}^{1}P(x)P(y)xy \: dx \:dy \\
        & =&\int_{0}^{1}P(x)x \: dx \int_{0}^{1}P(y)y \:dy \label{eq:ProbParejaInestable}
\end{eqnarray}

\noindent where $P(x)$ is the probability density of the energies of
        men and women in the Global Optimum. The integrals are nothing but the averages values of the energies of men and women, which are related to the the total energy of men $X$ and women $Y$ as $x=X/N$ and $y=Y/N$. In the fundamental state $X=Y=\frac{E^0}{2}=0.81\sqrt{N}$ then
\begin{eqnarray}
P_{h,m} &=& \frac{X}{N}\frac{Y}{N} \:\:=\:\: \left( \frac{E^0}{2N} \right) ^2 \nonumber \\
        &=& \frac{0.654}{N} \nonumber
\end{eqnarray}

\noindent $P_{m,w}$ is the probability that a generic couple $(m,w)$ is an unstable couple. Then, any man $m$ belongs to $N P(m,w)$ unstable couples, and, as there are $N$ men, the total number of unstable couples will be:

\begin{eqnarray}
N_{uc}  &=& N^2 P_{m,w} \nonumber \\
        &=& 0.654 N \nonumber \\
        &\propto& N \nonumber \\
\end{eqnarray}

\section{Acknowledgments}

We thank C. Trallero-Giner for useful discussions and comments. We
also acknowledge the support of the NET-61 from the ICTP.


\begin{thebibliography}{7}
\bibitem{Gus89} D. Gusfield and R.W. Irving, {\em The stable marriage
    problem}, MIT Press, Cambridge, MA, 1989
\bibitem{Gib92} R. Gibons, {\em A Primer in Game Theory}, Harvester
  Wheatsheaf, 1992
\bibitem{Mez85} M. Mezard and G. Parisi, J. Physique Lett {\bf 46}
  (1985) L771
\bibitem{Ome97} M.-J. Omero, M. Dzierzawa, M. Marsili and Y.-C. Zhang,
  J. Phys. I France {\bf 7} (1997) 1723 
\bibitem{Dzi00} M. Dzierzawa and M.-J. Omero, Physica A {\bf 287},
  (2000) 321 
\bibitem{Lau03} P. Laureti and Y.-C. Zhang, Physica A {\bf 324},(2003) 49 
\bibitem{Cal00} G. Caldarelli and A. Capocci, cond-mat/0008337
\bibitem{pepe} K. Huang, {\em Introduction to Statistical Physics},
  London, Taylor and Francis, 2001 
\end{thebibliography}
\end{document}